\documentclass[12pt]{article}

\usepackage{cite}
\usepackage{graphicx}
\usepackage{dcolumn}
\usepackage{amssymb}
\usepackage{amsmath}
\usepackage{epsfig}

\usepackage[colorlinks=true,urlcolor=blue,anchorcolor=blue,citecolor=blue,filecolor=blue,linkcolor=blue,menucolor=blue,pagecolor=blue]{hyperref}

\usepackage[small,bf]{caption}
\newcommand{\Htb}[1][]{\ensuremath{H^+\rightarrow t\,\bar{b}}}
\newcommand{\tbH}[1][]{\ensuremath{t\rightarrow b\,H^+}}

\newcommand{\tb}[1][]{\ensuremath{\tan^{#1}\!\beta}}

\newcommand{\mb}[1][]{\ensuremath{m_b^{#1}}}
\newcommand{\Dmb}[1][]{\ensuremath{\Delta\mb[#1]}}

\newcommand{\mg}[1][]{\ensuremath{M_{\tilde{g}}^{#1}}}
\newcommand{\msb}[2][]{\ensuremath{m_{\tilde{b}_{{#2}}}^{{#1}}}}
\newcommand{\mst}[2][]{\ensuremath{m_{\tilde{t}_{{#2}}}^{{#1}}}}

\newcommand{\sst}[1][]{\ensuremath{\sin^{#1}\!\theta_{\tilde{t}}}}
\newcommand{\cst}[1][]{\ensuremath{\cos^{#1}\!\theta_{\tilde{t}}}}
\newcommand{\ssb}[1][]{\ensuremath{\sin^{#1}\!\theta_{\tilde{b}}}}
\newcommand{\csb}[1][]{\ensuremath{\cos^{#1}\!\theta_{\tilde{b}}}}

\newcommand{\aS}[1][]{\ensuremath{\alpha_s^{#1}}}

\newcommand{\MS}[1][]{\ensuremath{\overline{\mathrm{MS}}}}

\newcommand{\ov}{\overline}
\newcommand{\lt}{\left}

\newcommand{\be}{\begin{equation}}
\newcommand{\ee}{\end{equation}}
\newcommand{\bc}{\begin{center}}
\newcommand{\ec}{\end{center}}

\newlength{\Oldarrayrulewidth}

\usepackage{amsmath}

\usepackage{dcolumn}
\usepackage{amssymb}
\usepackage{amsmath}
\usepackage{epsfig}

\def\lapp{\mathrel{\rlap{\raise.5ex\hbox{$<$}}
                    {\lower.5ex\hbox{$\sim$}}}}
\def\gapp{\mathrel{\rlap{\raise.5ex\hbox{$>$}}
                    {\lower.5ex\hbox{$\sim$}}}}
\newcommand{\gsim}{\lower.7ex\hbox{$\;\stackrel{\textstyle>}{\sim}\;$}}
\newcommand{\lsim}{\lower.7ex\hbox{$\;\stackrel{\textstyle<}{\sim}\;$}}
\begin{document}

\centerline {\bf SUSY scale uncertainties in  $m_b(m_b)$: $Br[B_s^0 \to \mu^+\mu^-]$ and cMSSM}
\vskip .5cm

\centerline{\sf Abhijit Samanta\footnote{The correspondence email address: abhijit.samanta@gmail.com}, Md Ansarul Haque}
\vskip .5cm
\centerline{\em Nuclear and Particle Physics Research Centre} 
\centerline{\em Department of Physics} 
\centerline{\em Jadavpur University}
\centerline{\em Kolkata 700 032, India}
\vskip .5cm
\begin{abstract}

We have studied  the SUSY scale dependence of supersymmetric radiative corrections 
to bottom quark mass on the branching ratio of $B_s^0 \to \mu^+\mu^-$ and  its impact on the cMSSM parameter space. 
The supersymmetric radiative corrections to bottom quark mass is in general evaluated at the SUSY scale, which is 
normally considered to be the geometric mean of stop masses. 
Then, the running of bottom quark mass is considered from the SUSY scale to the matching scale $M_W$ of the effective 
Hamiltonian which describes this decay. It is found that  the branching ratio varies drastically with the SUSY scale.  
Typically the $Br[B_s^0 \to \mu^+\mu^-]_{\rm untagged}$ varies $\sim 0.6 \times 10^{-9}$  due variation of bottom quark mass $\sim 35\%$ 
with the SUSY scale and these changes are very significant compared to their present uncertainties in the measurements.  
Moreover, these variations are very larger at the regions of parameter space which are favorable for producing  higgs mass 
around 125 GeV. Finally,  the confrontation of cMSSM with $B_s^0 \to \mu^+\mu^-$ is drastically relaxed if one lowers 
the SUSY scale and a significantly large parameter space becomes allowed in the phenomenologically interesting regions.   
\end{abstract}

\section{\bf Introduction} \label{Introduction}
Among the promising theories that have emerged over the past decades to solve the shortcomings of the standard model (SM) 
of particle physics, arguably the most beautiful and far reaching one is supersymmetry (SUSY). 
It represents a new type of symmetry that relates bosons and fermions, thus unifying forces (mediated by vector 
bosons) with matter (quarks and leptons), which endows space-time with extra fermionic dimensions. 
At present, from direct searches there is no evidence of SUSY particles at LHC, it excludes more and more parameter ranges 
depending on some specific conditions. From indirect searches, it also excludes parameter spaces of different SUSY models, even whole  model in some cases.
It is found that  the measured neutral higgs boson mass 
\cite{cmshiggs,atlashiggs} 
and the measured branching ratio $Br[B_s^0 \to \mu^+\mu^-]$  at CMS and LHCb \cite{CMS:2014xfa} are not simultaneously satisfied
except for a very small region of parameter space of the 
the constrained minimal supersymmetric standard model (cMSSM) \cite{Arbey:2012ax}.
In this work, we have studied  the SUSY scale dependence of supersymmetric radiative corrections 
to bottom quark mass on the branching ratio of $B_s^0 \to \mu^+\mu^-$ and  its impact on the cMSSM parameter space. 

Here, we consider the running of all parameters from $M_{\rm GUT}$ to SUSY scale
$M_{\rm SUSY}$, then the radiative corrections are calculated at $M_{\rm SUSY}$ and added to $m_b$ to find the $Y_b$.
Then, this $Y_b$ runs from $M_{\rm SUSY}$ to $M_W$ (the matching scale where effective Hamiltonian describing the decay is evaluated)
 using SM $\beta-$functions. 


The $M_{\rm SUSY}$ is considered to be the geometric mean of stop masses in literature  to minimize the 1-loop 
radiative corrections to Higgs potential. 
But, the 2-loop corrections normally have not the minima at geometric mean of stop masses.
There is no {strict theoretical reasoning} to consider $M_{\rm SUSY}$
to be fixed the geometric mean of stop masses. The SUSY is not considered spontaneously broken and
 the SUSY spectra may widely spread from $\sim 100$ GeV to a few TeV.
Here, it might be noted that SUSY scale uncertainty of radiative corrections 
to a physical quantity is not completely canceled by the running of the parameter through SM $\beta-$functions.

%
%

In this work we found that 
$Br[B_s^0 \to \mu^+\mu^-]_{\rm untagged}$ varies $\sim 0.6 \times 10^{-9}$  due variation of bottom quark mass $\sim 35\%$ 
with SUSY scale from $M_{\rm SUSY}=\sqrt{m_{\tilde t_L} m_{\tilde t_R}}$ to $M_W$  and these changes are very significant 
compared to their present uncertainties in the measurements.  
Moreover, these variations become larger for the regions of parameter space, which produces higgs mass around 125 GeV. 
Finally, we have shown that the confrontation of cMSSM with $B_s^0 \to \mu^+\mu^-$ is drastically relaxed if we lower 
the SUSY scale and a significantly large parameter space becomes allowed in the phenomenologically interesting region.


We organize the paper in the following way. We first briefly review the calculation of $Br[B_s^0 \to \mu^+\mu^-]$ in Section \ref{s:brbsmumu}, 
the supersymmetric radiative corrections $\Delta m_b$ in Section \ref{s:mbloop}, and
the impact of the SUSY scale dependence on the calculation of  $Br[B_s^0 \to \mu^+\mu^-]$ in Section \ref{s:scale}. Finally, we have shown
the change in allowed parameter space (APS) due to the change of SUSY  scale from   $\sqrt{m_{\tilde t_L} m_{\tilde t_R}}$ to $M_Z$
in Section \ref{s:cmssm}.

\section{Calculation of $Br[B_s^0 \to \mu^+\mu^-]$  in MSSM }\label{s:brbsmumu}

The effective Hamiltonian describing the $b \to s \ell^+ \ell^-$ transitions has the following generic structure:
\begin{equation}
        H = \frac{G_F}{\sqrt{2}} 
        \frac{\alpha_{EM}}{2\pi \sin^2\theta_W} \xi_t
        \left[ C_S Q_S + C_P Q_P + C_A Q_A \right].
        \label{hami}
\end{equation}
Here $G_F$ is the Fermi constant, $\alpha_{EM}$ is the electromagnetic
fine structure
constant and $\theta_W$ is the Weinberg angle. The CKM elements are
contained in $\xi_t= V_{tb}^* V_{td^{\prime}}$.  The operators in
Eq. \ref{hami} are
\begin{equation}
        Q_S \; = \; m_b \, \bar b P_L d^{\prime} \, \bar \ell \ell,  \qquad
        Q_P \; = \; m_b \, \bar b P_L d^{\prime} \, \bar \ell \gamma_5 \ell,
                \qquad  
        Q_A \; = \; \bar b \gamma^{\mu} P_L d^{\prime} \, 
                \bar \ell \gamma_{\mu} \gamma_5 \ell,
        \label{ops}
\end{equation}
where $P_L = (1-\gamma_5)/2$ is the left-handed projection operator.

\begin{eqnarray}
  \label{eq:Bsmm_formula}
\mathrm{BR}(B_s\to\mu^+\mu^-)&=&\frac{G_F^2 \alpha^2}{64\pi^3}f_{B_s}^2
m_{B_s}^3 |V_{tb}V_{ts}^*|^2\tau_{B_s}\sqrt{1-\frac{4m_\mu^2}{m_{B_s}^2}}\\
&&\times\left\{\left(1-\frac{4m_\mu^2}{m_{B_s}^2}\right)
  |C_S|^2+\left|(C_P-2C_A\frac{m_\mu}{m_{B_s}}\right|^2\right\}\,,\nonumber  
\end{eqnarray}
where $f_{B_s}$ is the $B_s$ decay constant, $m_{B_s}$ is the $B_s$ meson mass and $\tau_{B_s}$ is its mean lifetime. 

The one-loop corrected Wilson coefficients $C_{S,P}$ are taken from \cite{Bobeth:2001sq} and $C_A$ from \cite{Logan:2000iv}.  The Wilson coefficients $C_{S,P}$ 
have been multiplied by $1/{(1+\epsilon_b\tan\beta)}^2$, where $\epsilon_b$ incorporates the full supersymmetric one-loop corrections to the bottom 
Yukawa coupling. 

\subsection{Supersymmetric one loop correction $\Delta m_b$}\label{s:mbloop}

\begin{figure*}[htb]
\vspace*{-2cm}
\includegraphics[height=17.5cm]{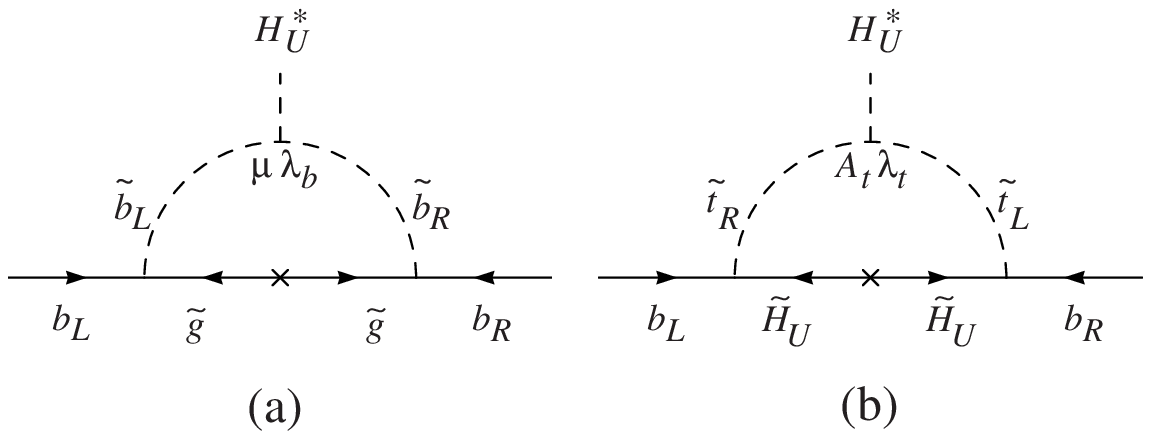}
\vspace*{-9cm}
\caption{One loop SUSY-QCD and SUSY-EW diagrams contributing to $b$ quark mass.}
\label{f:loop-H2}
\end{figure*}

The tree level $H_2^0 b \bar b$ is forbidden in MSSM. But, a non-vanishing  correction to bottom quark Yukawa coupling $\Delta h_b$ 
can be generated at one loop level by the diagrams in Fig. \ref{f:loop-H2} \cite{Carena:1999py}. 
Once the higgs fields $H_{1,2}^0$ acquire vacuum expectation values
$v_{1,2},$ the tree level relation between bottom mass $m_b$ and Yukawa coupling $h_b$ is shifted by 

\begin{eqnarray}
  m_b = h_b v_1 \longrightarrow m_b = v_1 ( h_b + \Delta h_b \tan\beta) = h_b v_1 (1+ \Delta m_b).
\end{eqnarray}
Then, 
\begin{eqnarray}
   h_b = \frac{m_b}{v} \frac{1}{1+\Delta m_b} \tan\beta,
\label{eq:hb-mb}
\end{eqnarray} 
where, $v=\sqrt{v_1^2+v_2^2}$, $\tan\beta= \frac{v_2}{v_1}$ and 
  $ \epsilon_b\tan\beta = \Delta m_b = \frac{\delta m_b}{m_b}$.
The $\Delta h_b$ is loop-induced Yukawa coupling connected with supersymmetric QCD and electroweak corrections and
\begin{eqnarray}
 \Delta m_b = \frac{\Delta h_b}{h_b} \tan\beta = \Delta m_b^{\rm SQCD} + \Delta m_b^{\rm SEW}.  
\end{eqnarray}
 The supersymmetric QCD
corrections of fig.~\ref{f:loop-H2} is given by \cite{copw}
\begin{eqnarray}
  \label{eq:dmbSQCD}
  \Dmb[SQCD] &=& \frac{2\aS}{3\pi}\mg 
                        \mu\tb\,I(\msb{1},\msb{2},\mg)\,.
\end{eqnarray}
Here, \aS\ is the strong coupling constant and $\mu$ is the mass parameter
coefficient of the $\epsilon_{ij} H_i^1 H_j^2$ term in the superpotential.
The vertex function $I$, depends on the masses \msb{1,2} of the two
bottom squark mass eigenstates and the gluino mass \mg, reads \cite{copw}
\begin{equation}
  \label{eq:I}
  I(a,b,c)=\frac{1}{(a^2-b^2)(b^2-c^2)(a^2-c^2)}
  \left(a^2b^2\log{\frac{a^2}{b^2}}
        +b^2c^2\log{\frac{b^2}{c^2}}
        +c^2a^2\log{\frac{c^2}{a^2}}\right)\,.
\end{equation}
The electroweak correction reads \cite{pierce}
\begin{eqnarray}
  \label{eq:dmbSEW}
  \Dmb[SEW] &=&
  \frac{h_t^2}{16\pi^2}\,\mu A_t\tb\,I(\mst{1},\mst{2},\mu)\nonumber\\
  &-& \frac{g^2}{16\pi^2}\,\mu M_2\tb
    \left[\vphantom{\frac{1}{2}}
            \cst[2]\,I(\mst{1},M_2,\mu)
           +\sst[2]\,I(\mst{2},M_2,\mu)\right.\nonumber\\
  &&\phantom{\frac{g^2}{16\pi^2}\,\mu\tb M_2}
+\frac{1}{2}\csb[2]\,I(\msb{1},M_2,\mu)
+\frac{1}{2}\left.\ssb[2]\,I(\msb{2},M_2,\mu)\right] \,.
\end{eqnarray}
                      
If we consider the approach of effective field theory, the heavy particles, squarks and gluinos are integrated out 
and the interaction mediated by the loop diagrams  of Fig. \ref{f:loop-H2} is represented by effective operator $\bar b b H_2^0$. Its Wilson 
coefficient equals to $-\Delta h_b$ $(Q= Q_0 = M_{\rm SUSY})$,  but the relation  between $h_b$ and     $m_b$ (Eq. \ref{eq:hb-mb}) is defined at scale $Q=m_b$.
Since we encounter same operator $\bar bb$ in tree level as well as in leading order, the renormalization group running down to $m_b$ are 
identical for both cases.     
The running Yukawa coupling $\overline{h_b} (Q=m_b)$  in the desired relation becomes
\begin{eqnarray}
   \ov{h}_b \lt(Q=\mb \right) & = &   
 \frac{\ov{m}_b \lt( Q=\mb \right)}{v} \, 
 \frac{1}{1+\Delta\mb \lt( Q=Q_{0} \right)}\,\tb.
\label{eq:uli:wb}
\end{eqnarray}
The   overlined quantities are in $\overline{MS}$ scheme. 
The details of the two loop supersymmetric QCD corrections, top induced SUSY electroweak corrections, including NNLO corrections can be found in \cite{Ciuchini:1997xe,Degrassi:2000qf,Noth:2008tw,Noth:2010jy,Mihaila:2010mp,Ghezzi:2017enb}. 
Now, it is very important to see whether there is any dependence on the scale $Q_{0}$ or not for  determination of $h_b$ at any value of $Q$
below $Q_0$.

\begin{figure*}[htb]
\bc
\includegraphics[width=9cm,angle=0]{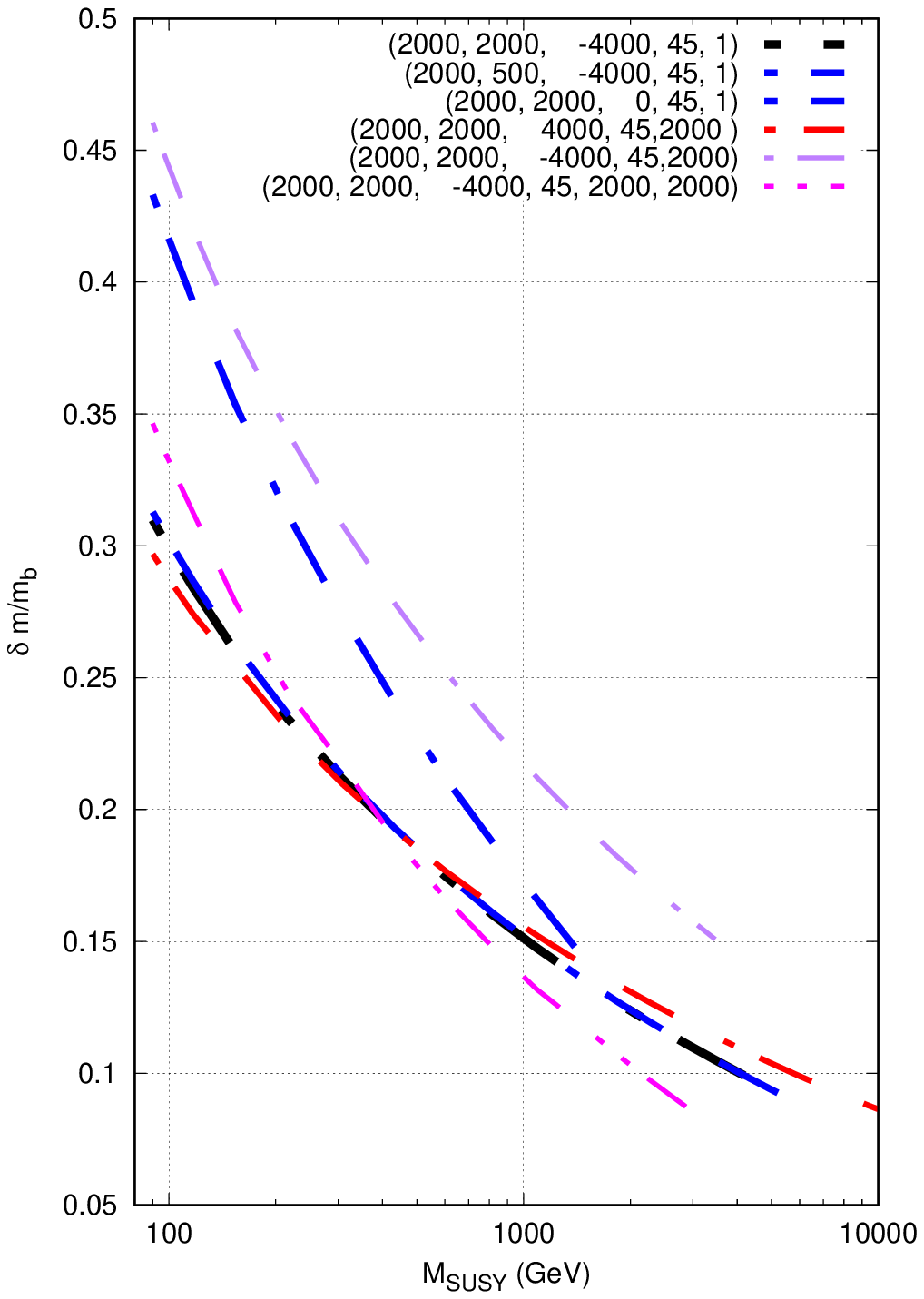}
\ec
\caption{The  variation of $\frac{\Delta m_b}{m_b}$ as a function of $M_{SUSY}$ for different sets of inputs in the format ($m_0, m_{1/2}, A_0, \tan\beta$) for cMSSM with $\mu>0$, ($m_0, m_{1/2}, A_0, \tan\beta$, $\mu$) for NUHM1, and ($m_0, m_{1/2}, A_0, \tan\beta$, $\mu$, $m_A$) for NUHM2. }
\label{f:q0delmb}
\end{figure*}

The effective theory describing the $b$ decays is obtained using Operator Product Expansion (OPE) at  the matching scale $M_W$.
%
The value of $Q_0(=M_{\rm SUSY})$ is normally chosen to be the geometric mean of stop masses in literature \cite{superiso,microomega}.   
The $\Delta m_b$  depends directly on  $h_t$, $\alpha_s$, $A_t$ (see Eq. \ref{eq:dmbSQCD} and \ref{eq:dmbSEW}) and they  drastically 
vary with renormalization group evolution (RGE) scale. This leads to  a significant variation
in $m_b(m_b)$ with  $Q_0$  through the evaluation of $\Delta m_b$. We have plotted $\Delta m_b/m_b$ vs. $Q_0$ in Fig. 
\ref{f:q0delmb}. We find that $\Delta m_b/m_b$ varies 30\% to 45\% when we vary the $M_{\rm SUSY}$ from 
$\sqrt{m_{\tilde t_L} m_{\tilde t_R}}$ to $M_W$ depending on the cMSSM input parameters.

\section{Result}
\subsection{Impact of SUSY scale dependence in evaluation of $m_b(m_b)$ on $Br[B_s^0 \to \mu^+\mu^-]$}\label{s:scale}
In calculation of $Br[B_s^0 \to \mu^+\mu^-]$ the factor
$(1+\epsilon_b\tan\beta)^{-2}$ is multiplied with $C_{S,P}$ to incorporate the supersymmetric radiative corrections $\Delta m_b$ to bottom quark mass.
In $\epsilon_b$, the contribution from gluino-exchange diagram is directly proportional to
strong coupling constant  $\alpha_s$ (see Eq. \ref{eq:dmbSQCD}) and contribution from higgsino-exchange diagram is directly proportional to 
top Yukawa coupling $Y_t$ (see Eq. \ref{eq:dmbSEW}). Both $\alpha_s$ and $Y_t$ are strongly RGE scale dependent
and their values increase very rapidly with the decrease in RGE scale. 
This trend is generic feature  for all SUSY GUT models 
and hence the variation of $\Delta m_b$ with the SUSY scale for all SUSY GUT models is dominantly controlled by $\alpha_s$ and $Y_t$. 

\begin{figure*}[htb]
\bc\includegraphics[width=9cm,angle=0]{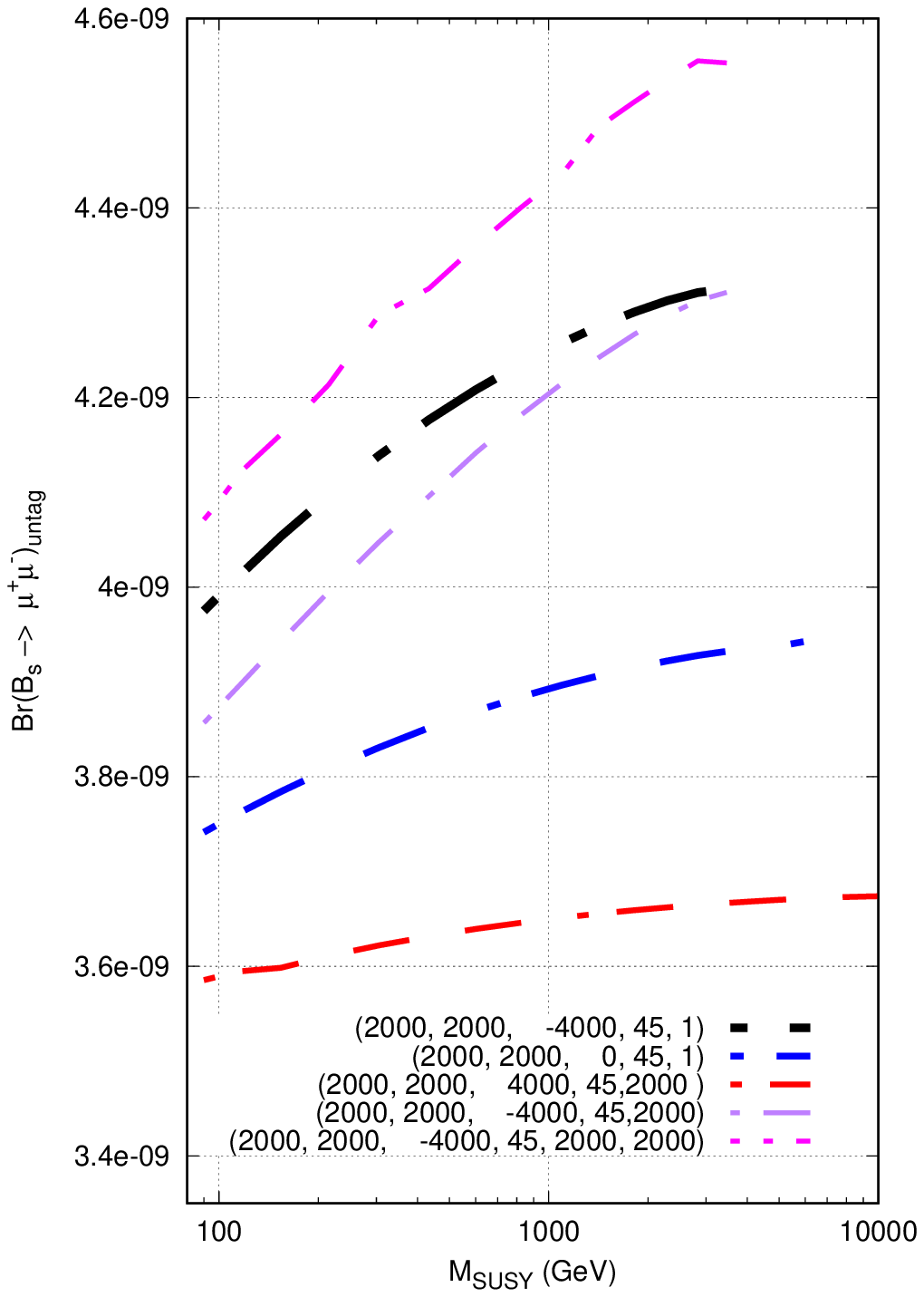}
\ec
\caption{The variation for $Br[B_s^0 \to \mu^+\mu^-]_{\rm untag}$ as a function of $M_{SUSY}$ for different sets of inputs in the format ($m_0, m_{1/2}, A_0, \tan\beta$) for cMSSM with $\mu>0$, ($m_0, m_{1/2}, A_0, \tan\beta$, $\mu$) for NUHM1, and ($m_0, m_{1/2}, A_0, \tan\beta$, $\mu$, $m_A$) for NUHM2. }
\label{f:q0bs2mumu}
\end{figure*}

The program package SuSpect \cite{suspect} is used to generate supersymmetric spectra  and the package  
 SuperIso \cite{superiso} is used for calculation of ${Br}[B_s^0 \rightarrow \mu^+\mu^-]$.
The variations of $Br[B_s^0 \to \mu^+\mu^-]$ with $Q_0$ 
are shown in Fig. \ref{f:q0bs2mumu}. We see that $Br[B_s^0 \to \mu^+\mu^-]$ can change up to $\simeq 10\%$. 

The measured value  of ${Br}[B_s^0 \rightarrow \mu^+\mu^-]$  at LHC is $\left(3.0\pm 0.6^{\,+0.3}_{\,-0.2}\right)\times 10^{-9}$ \cite{Aaij:2017vad}.
The results have been demonstrated
 for cMSSM, NUHM1, NUHM2, but it happens to have similar trends for  all SUSY GUTs for the reasons explained earlier.
The scale dependence is relatively stronger for larger negative $A_0$ with larger $\tan\beta$ and relatively weaker for positive $A_0$.
It should be noted here that these larger negative values of $A_0$ with larger values of $\tan\beta$ are also favourable for producing higgs 
mass around 125 GeV \cite{Buchmueller:2013psa,Baer:2011ab,Arbey:2012ax}. 


\begin{figure*}[htb]
\bc
\includegraphics[width=6cm,angle=0]{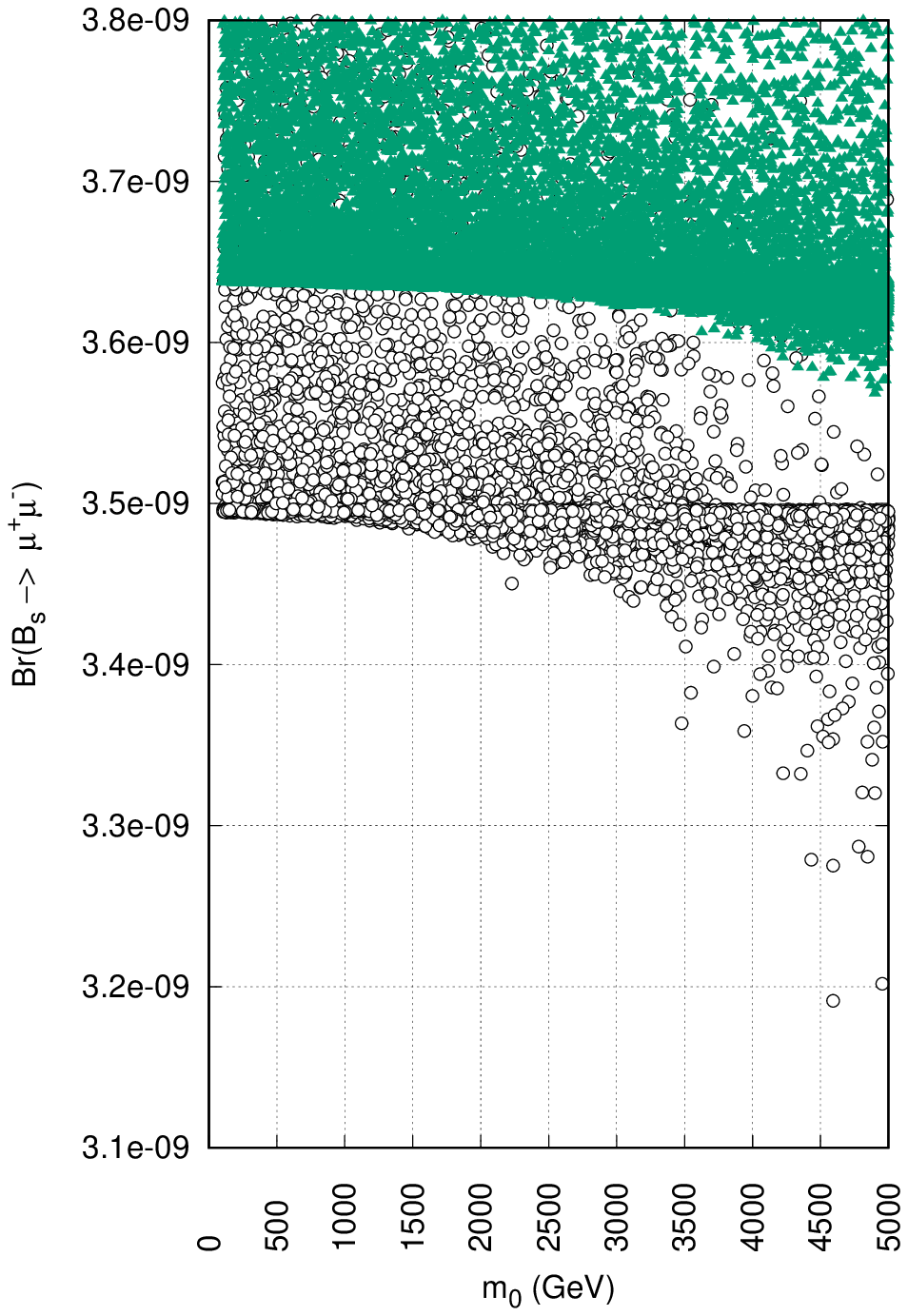}
\includegraphics[width=6cm,angle=0]{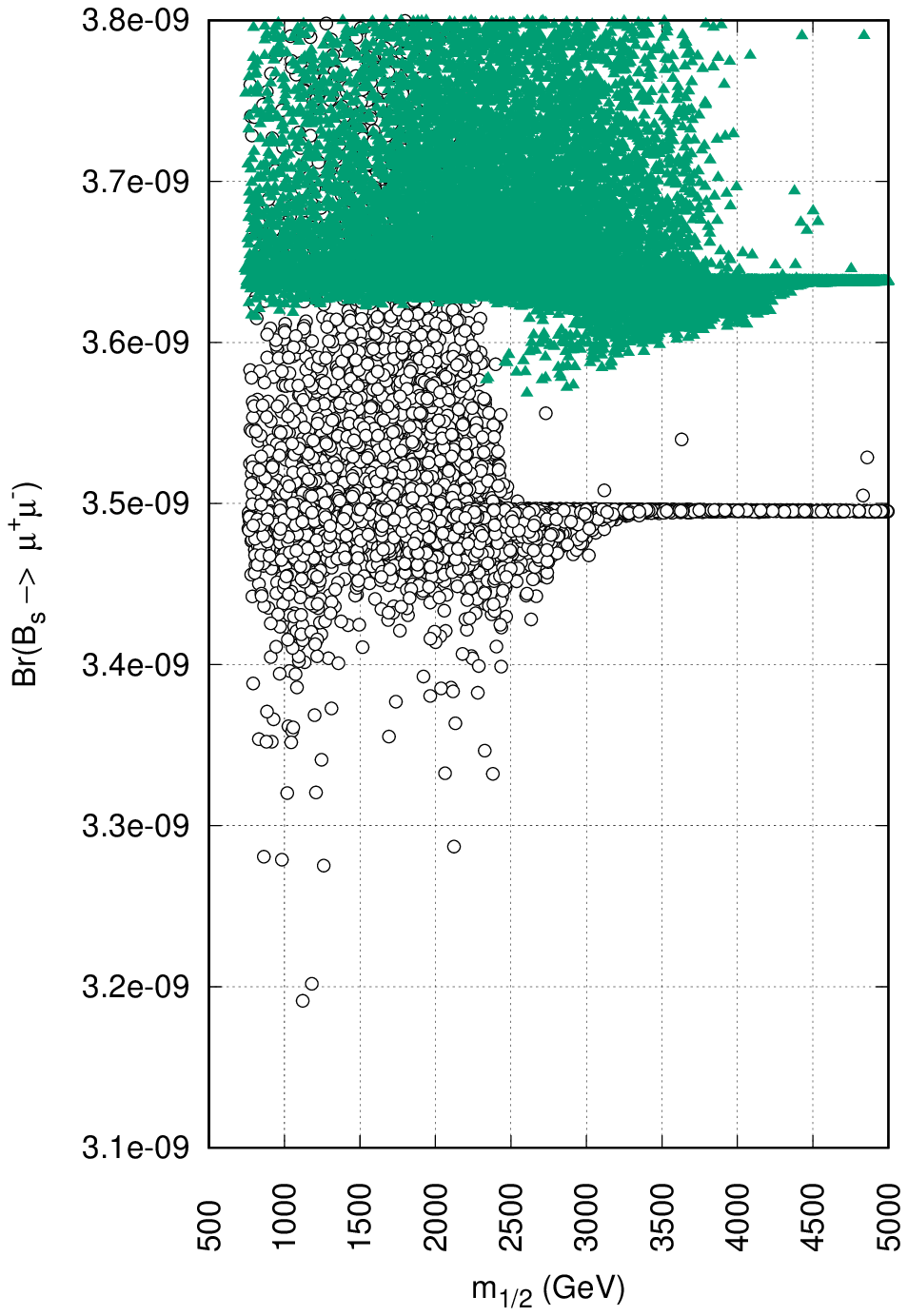}
\includegraphics[width=6cm,angle=0]{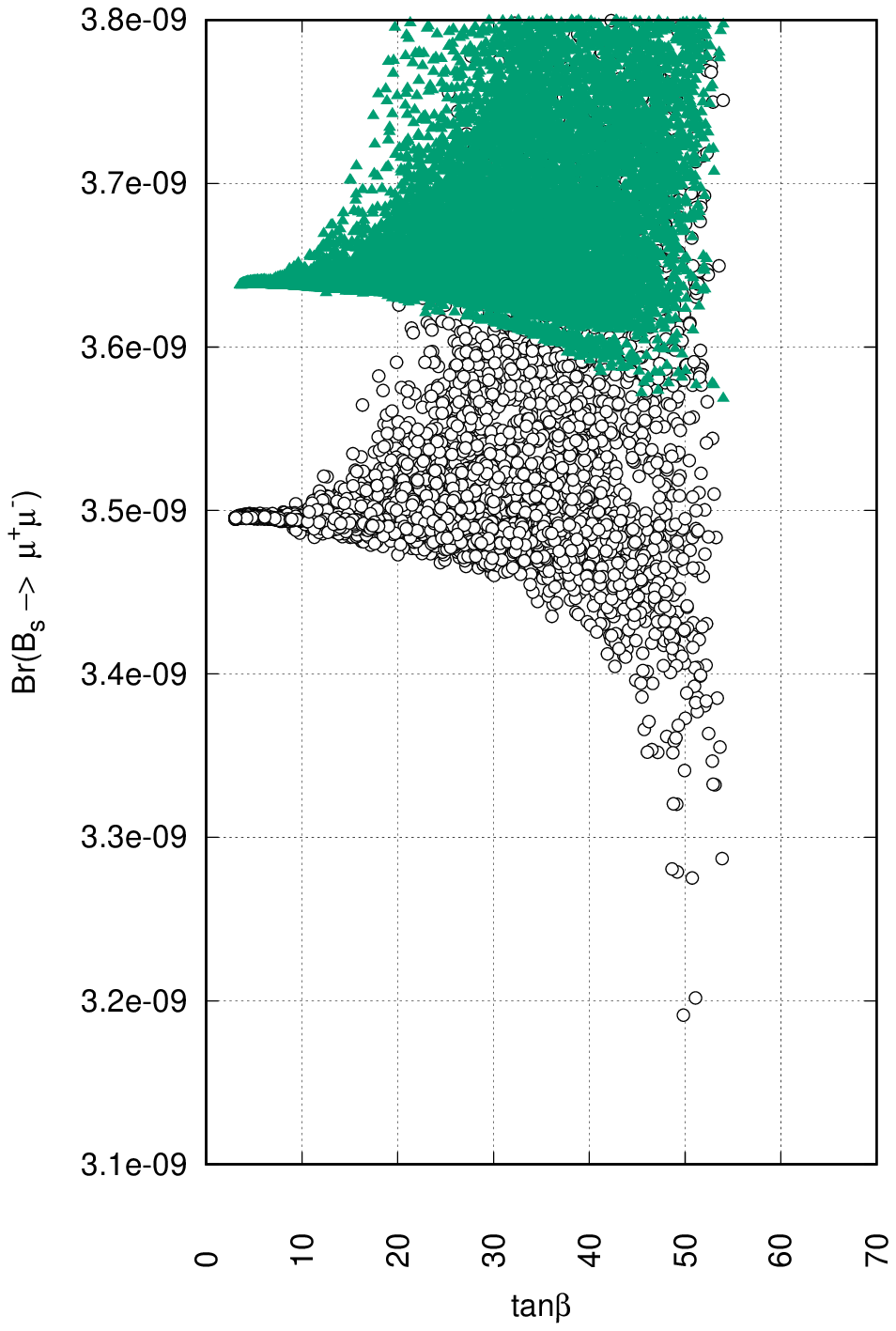}
\includegraphics[width=6cm,angle=0]{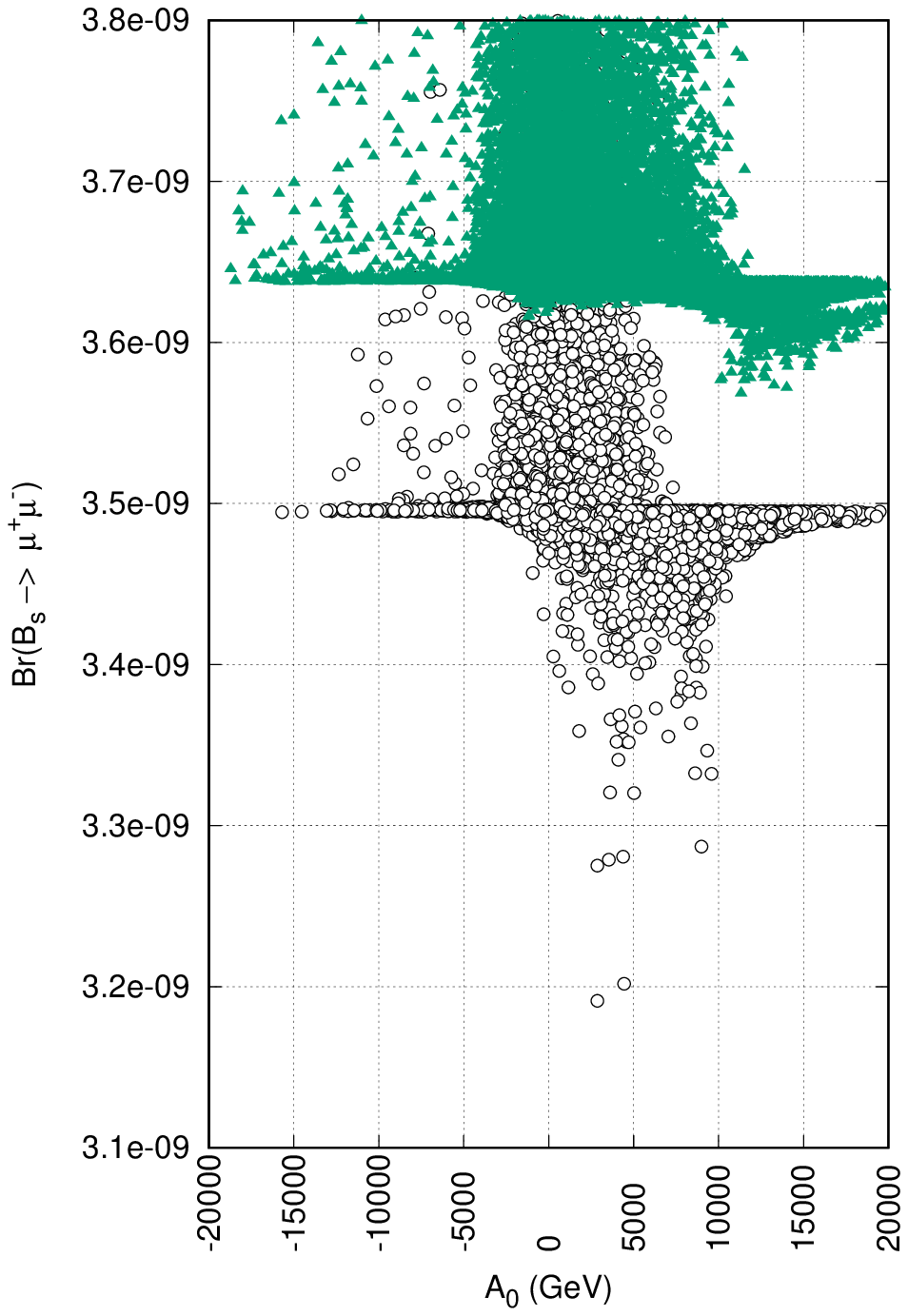}
\ec
\caption{The untagged $Br[B_s^0 \to \mu^+\mu^-]$ vs. cMSSM parameters $m_0$, $m_{1/2}$, $A_0$, and $\tan\beta$ in agreement
with experimental mass bounds as discussed in the text. The points denoted by triangle evaluated at  $M_{\rm SUSY}=\sqrt{m_{\tilde t_L} m_{\tilde t_R}}$ 
circles at  $M_{\rm SUSY}=M_Z.$} 
\label{f:bscmssm}
\end{figure*}

\begin{figure*}[htb]
\bc
\includegraphics[height=6.0cm, angle=0]{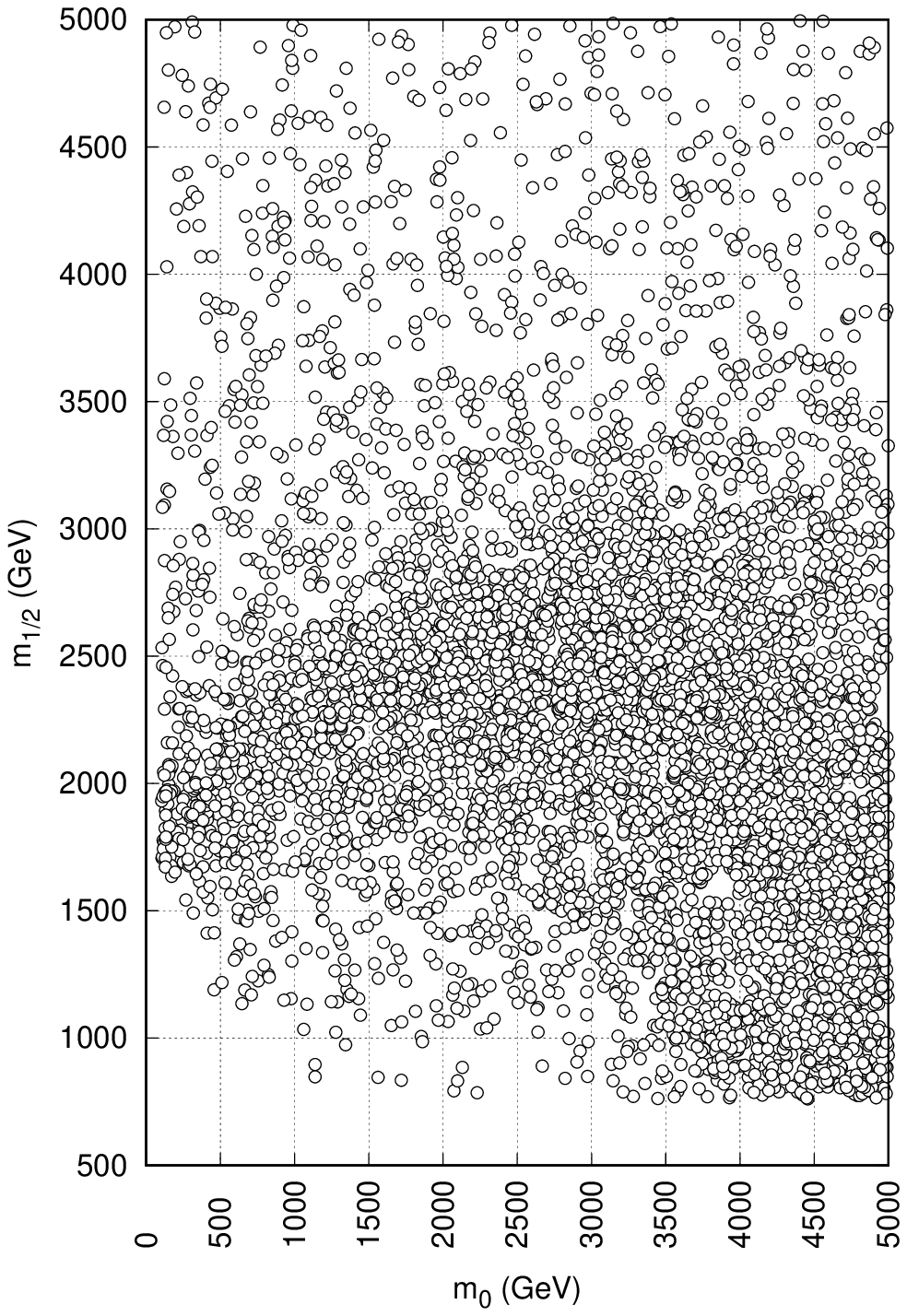}
\includegraphics[height=6.0cm, angle=0]{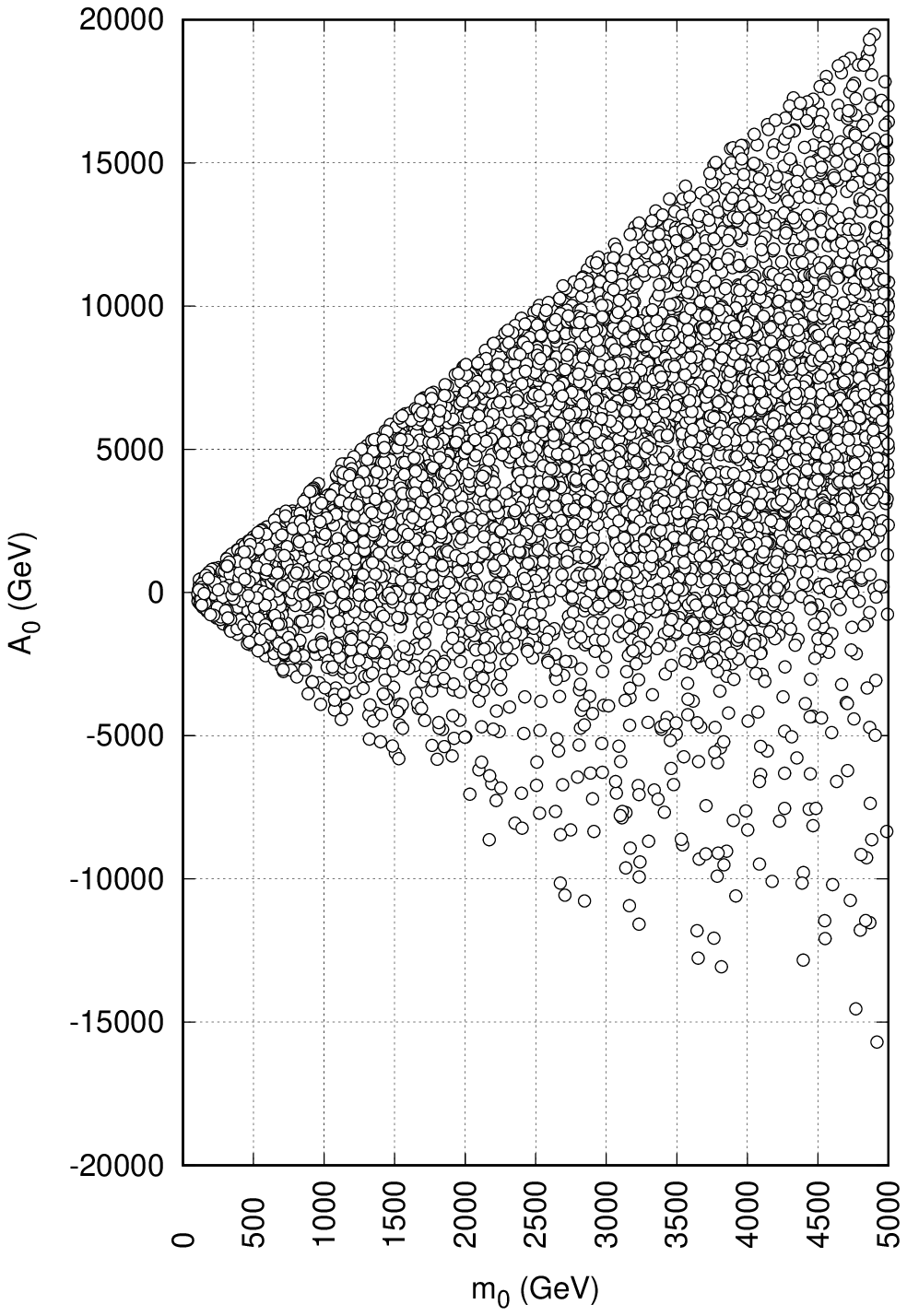}
\includegraphics[height=6.0cm, angle=0]{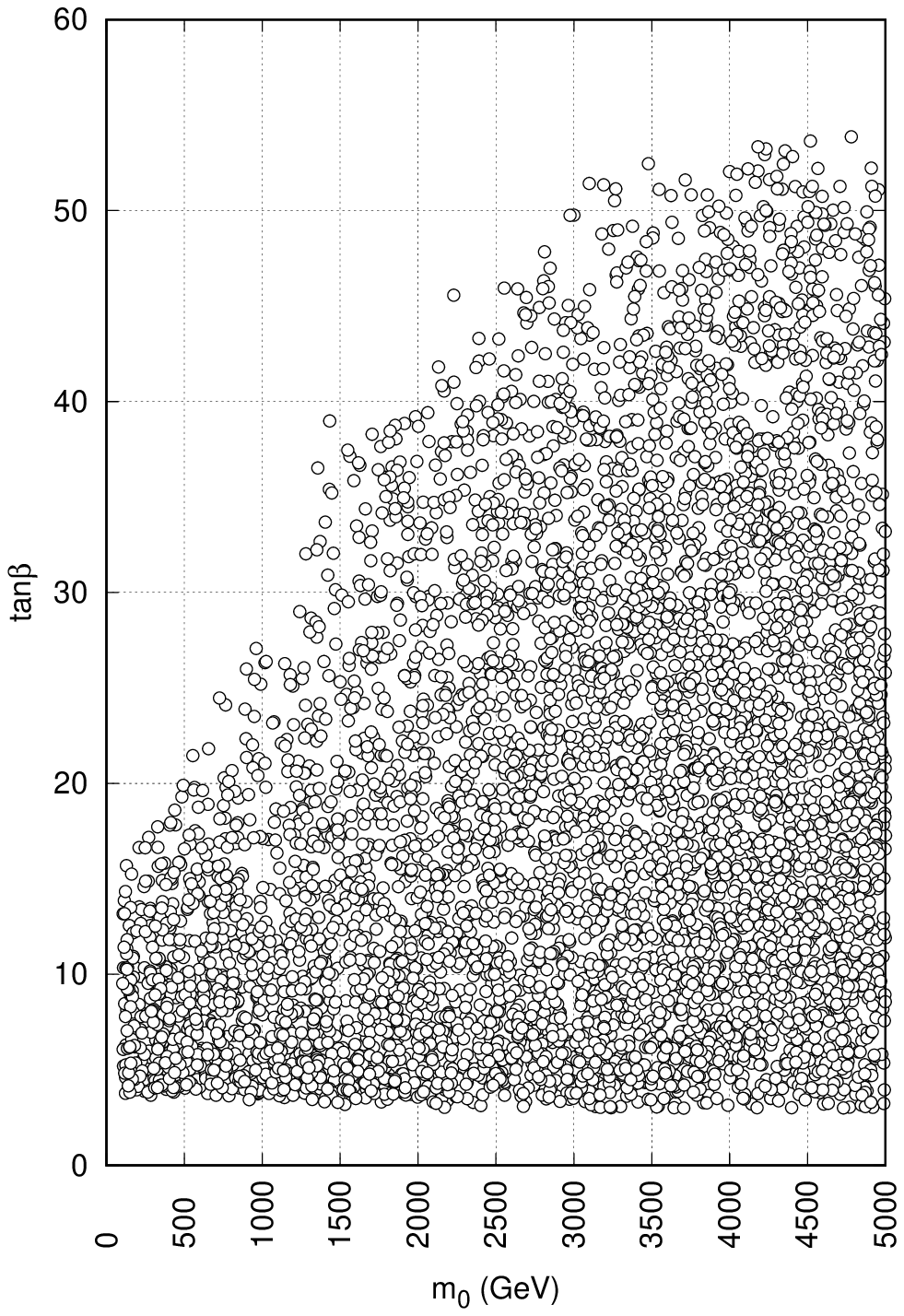}
\includegraphics[height=6.0cm, angle=0]{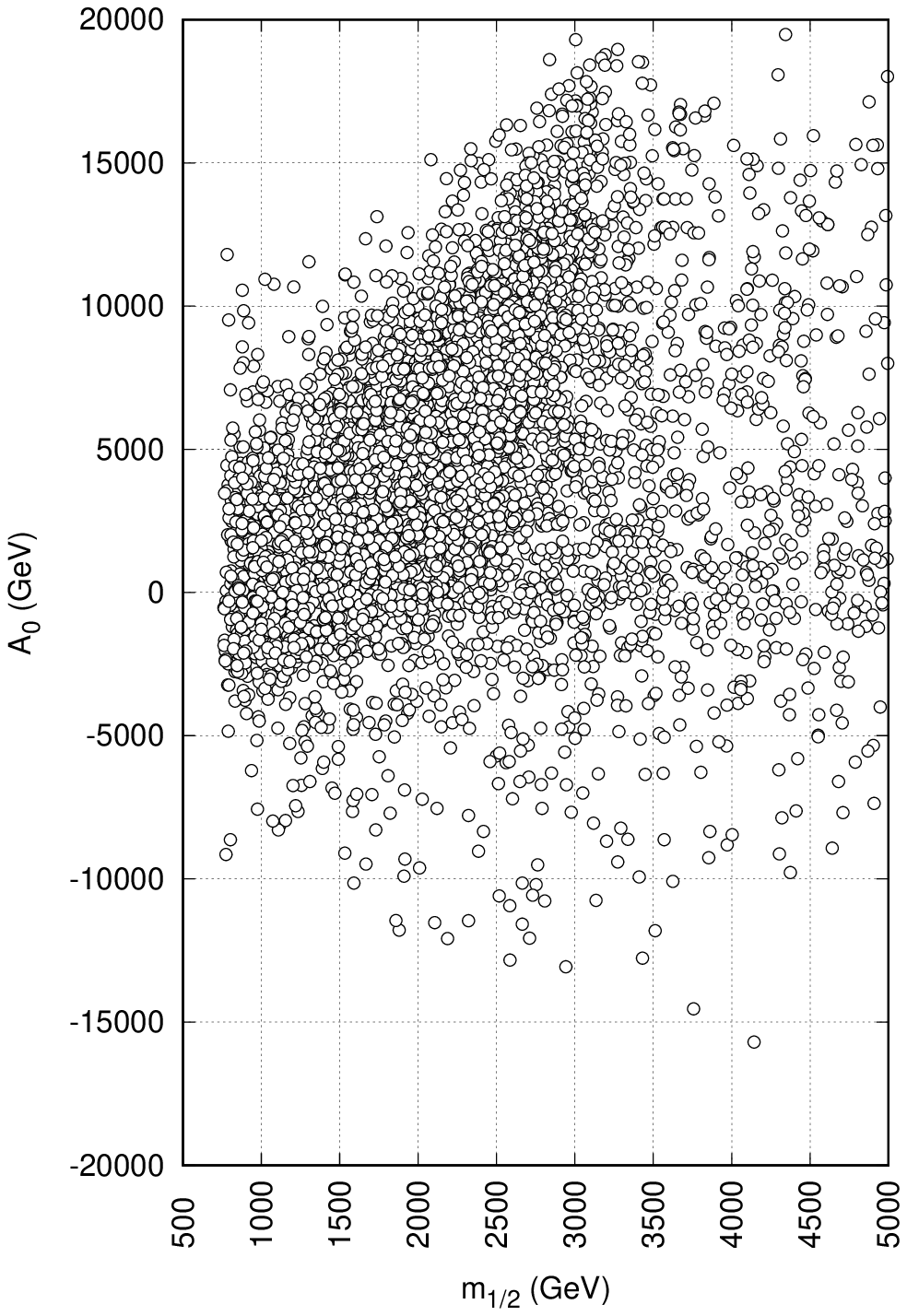}
\includegraphics[height=6.0cm, angle=0]{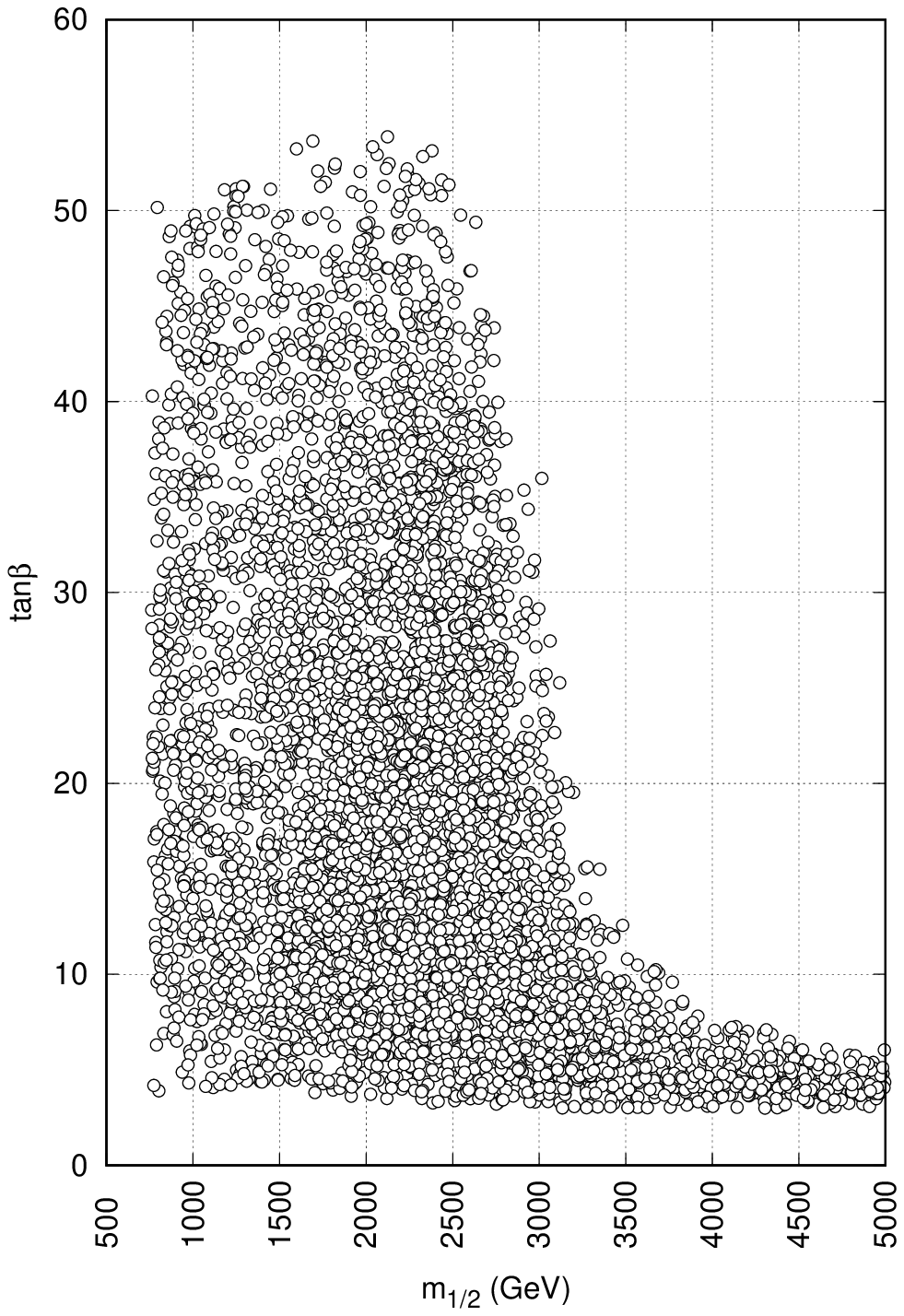}
\includegraphics[height=6.0cm, angle=0]{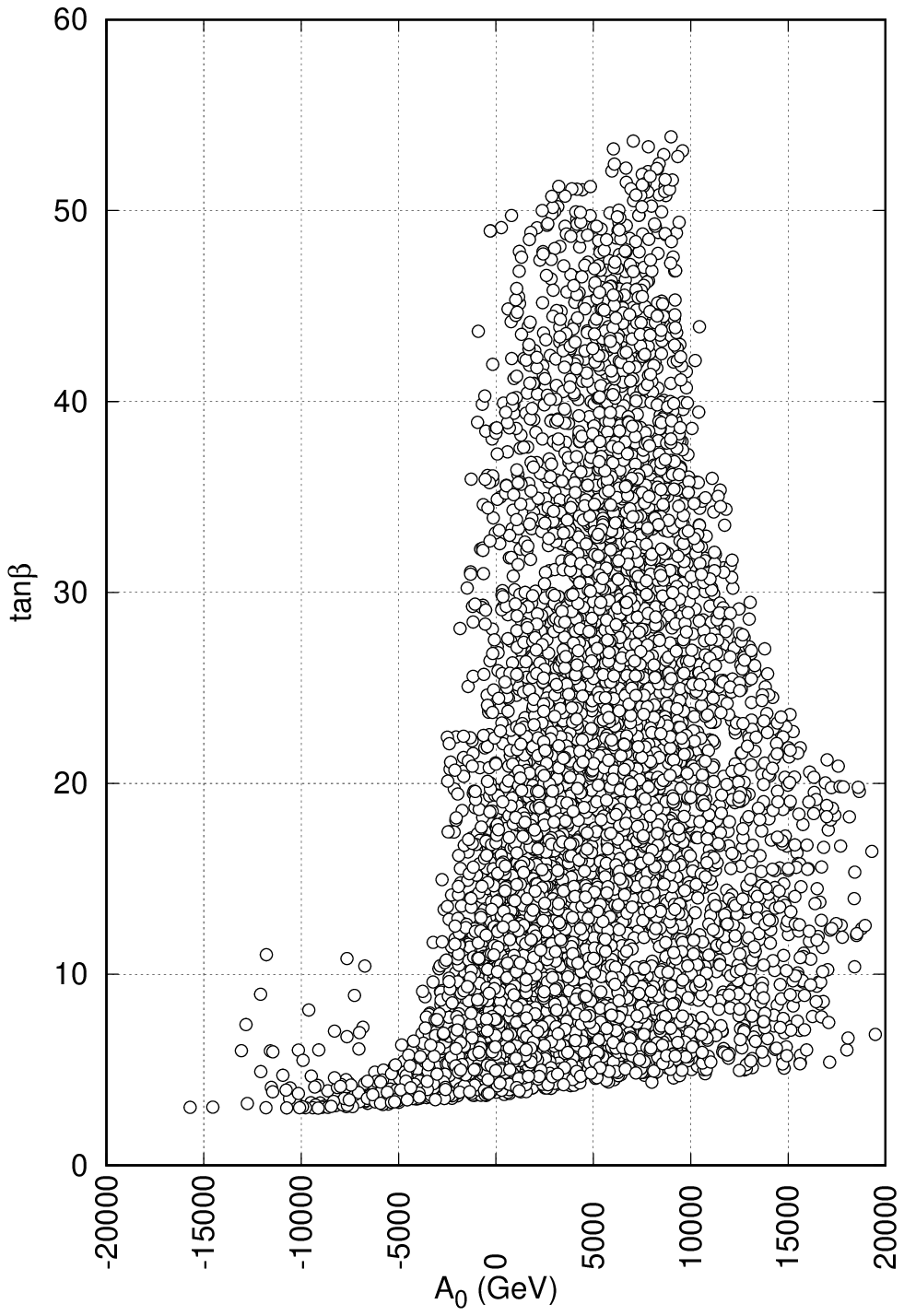}
\ec
\caption{The allowed cMSSM parameter space  in $m_0-m_{1/2}$, $m_{0}-A_0$, $m_0-\tan\beta$,  $m_{1/2}-A_0$, $m_{1/2}-\tan\beta$, and $A_0-\tan\beta$ planes, respectively. The $Br[B_s^0 \to \mu^+\mu^-]_{\rm untag} < 3.5 \times 10^{-9}$ along with experimental mass bounds as discussed in the text are considered.
}
\label{f:cmssm}
\end{figure*}
\subsection{ The cMSSM parameter space}\label{s:cmssm}
We have scanned the cMSSM parameter space for  100 GeV $\le m_0 \le$ 5000 GeV, 100 GeV $\le m_{1/2} \le$ 5000 GeV, 
$-4\le A_0/m_0\le +4$, $3\le\tan\beta\le 60$, and sign$(\mu) = +1$ (which is in better agreement with muon 
$(g-2)$ constraint \cite{Chattopadhyay:2001vx}) and calculated the untagged $Br[B_s^0 \to \mu^+\mu^-]$, along with sparticle masses using 
both $M_{\rm SUSY} = \sqrt{m_{\tilde t_L} m_{\tilde t_R}}$ and $M_{\rm SUSY} =M_Z$. 
It is to be noted here that we have  considered $Q_0 = M_{\rm SUSY}$ to be the EWSB scale as generally considered
in the literature.

In Fig. \ref{f:bscmssm} we  compare the $[B_s^0 \to \mu^+\mu^-]_{\rm untag}$ as a function of 
$m_0$, $m_{1/2}$, $A_0$, and $\tan\beta$ for both of these SUSY scales. 
For both cases, we consider 124.5GeV $<m_h<$ 125.5GeV, and mass bounds  
$m_{\tilde t_1} > 363$ GeV,  $m_{\tilde \chi_1^0} > 328$ GeV, $m_{\tilde g}> 1820$ GeV \cite{Han:2016gvr}, and 
$\tan\beta$ dependent mass bound on CP-odd scalar 
higgs mass $m_{A^0}$ \cite{ATLAS:2016fpj}. We should note here that there will be very little effect on the allowed 
parameter space with the change of experimental bounds since we present scattered plots and inputs are chosen randomly. 
It shows 
 $Br[B_s^0 \to \mu^+\mu^-]_{\rm untag}$ is significantly lowered for $M_{\rm SUSY} =M_Z$ for the reason discussed in Section \ref{s:scale},
which significantly weakens the confrontation of cMSSM with $B_s^0 \to \mu^+\mu^-$.

For providing an idea about the sparticle masses 
we plot the  cMSSM parameter space  in Fig. \ref{f:cmssm}  in $m_0-m_{1/2}$, $m_{0}-A_0$, $m_0-\tan\beta$,  $m_{1/2}-A_0$, 
$m_{1/2}-\tan\beta$, and $A_0-\tan\beta$ planes, respectively. Here, we consider an additional constraint 
$Br[B_s^0 \to \mu^+\mu^-]_{\rm untag} < 3.5\times 10^{-9}$  along with the mass bounds as discussed earlier. 
We find that no point is allowed for $Q_0=M_{\rm SUSY}=\sqrt{m_{\tilde t_L} m_{\tilde t_R}}$. This confrontation 
has been reported in \cite{Arbey:2012ax}. 
But, we find that a large parameter space becomes  allowed for $M_{\rm SUSY} =M_Z$ in the phenomenologically 
interesting regions.

We also note here that there are uncertainties in $m_h$ calculations from one program package to another 
due to considering or not considering of different NLO and NNLO corrections and the value of 
$m_h$ also varies very significantly with very slight change in SM input parameters.  Our focus is only to see 
the general trend in the change of APS with $M_{\rm SUSY}$ and this trend should not alter due to 
uncertainties in $m_h$ calculations. 
This can be clarified from the Table 1 of reference \cite{Samanta:2014gla}.
In \cite{Samanta:2014gla}, it has been shown that  how $m_h$ increases in  cMSSM with decrease in
electroweak symmetry breaking scale from $M_{\rm SUSY}=\sqrt{m_{\tilde t_L} m_{\tilde t_R}}$ to $M_Z$  
and how it leads to  changes in the APS of cMSSM parameter space. 

\section{Conclusion}
It is seen  that $B_s^0 \to \mu^+\mu^-$ varies $\sim 0.6 \times 10^{-9}$  due to variation of bottom quark mass $\sim 35\%$ with the 
SUSY scale and these changes are very significant compared to its present uncertainty in its measurement. 
These variations are seen to be significant for the regions of parameter space of   SUSY models, which 
produces  higgs mass around 125 GeV.  We show that confrontation of cMSSM with $B_s^0 \to \mu^+\mu^-$ is relaxed 
drastically if we lower the SUSY scale and a large region of parameter space becomes allowed in phenomenologically interesting region.

{\it \bf Acknowledgments:} AS acknowledges the support from Scientific and Engineering Research Board, 
Department of Science and Technology, Govt. of India through the research grant 
SB/S2/HEP-003/2013 and the use of computational facility under DST-FIST program. AH acknowledges the support
of CSIR, India for his fellowship.

\end{document}